\documentclass[twocolumn,iop]{emulateapj}
\usepackage{apjfonts}
\usepackage{color} 

\usepackage{amsmath,graphicx,longtable,hyperref}
\usepackage{natbib,threeparttable,rotating}
\usepackage{ulem}

\newcommand{\hbeta}{H{$\beta$}}
\newcommand{\halpha}{H{$\alpha$}}

\newcommand{\CIV}{C{\sevenrm IV}}

\def\FeII{Fe\,{\sc ii}}

\def \OIII {[O\,{\sc iii}]}

\newcommand{\bracket}[1]{\left\langle#1\right\rangle}
   \font\sevenrm=cmr7 scaled 1000
\newcommand{\comments}[1]{}

\def\kms{{\rm km\,s^{-1}}}
\def\Rfe{R_{\rm FeII}}

\slugcomment{ApJL in press}

\begin{document}

\title{Dissecting the quasar main sequence: insight from host galaxy properties}

\author{Jiayi Sun$^{1,2}$ and Yue Shen$^{2,3,4}$}

\altaffiltext{1}{Tsinghua Center for Astrophysics, Department of Physics, Tsinghua University, Beijing 100084, China}
\altaffiltext{2}{Carnegie Observatories, 813 Santa Barbara Street, Pasadena,
CA 91101, USA}
\altaffiltext{3}{Kavli Institute for Astronomy and Astrophysics, Peking University, Beijing 100871, China}
\altaffiltext{4}{Hubble Fellow}

\shorttitle{DISSECTING THE QUASAR MAIN SEQUENCE}
\shortauthors{SUN \& SHEN}

\begin{abstract}
The diverse properties of broad-line quasars appear to follow a well-defined main sequence along which the optical \FeII\ strength increases. It has been suggested that this sequence is mainly driven by the Eddington ratio ($L/L_{\rm Edd}$) of the black hole (BH) accretion. Shen \& Ho demonstrated with quasar clustering analysis that the average BH mass decreases with increasing \FeII\ strength when quasar luminosity is fixed, consistent with this suggestion. Here we perform an independent test by measuring the stellar velocity dispersion $\sigma_*$ (hence the BH mass via the $M-\sigma_*$ relation) from decomposed host spectra in low-redshift Sloan Digital Sky Survey quasars. We found that at fixed quasar luminosity, $\sigma_*$ systematically decreases with increasing \FeII\ strength, confirming that Eddington ratio increases with \FeII\ strength. We also found that at fixed luminosity and \FeII\ strength, there is little dependence of $\sigma_*$ on the broad \hbeta\ FWHM. These new results reinforce the framework put forward by Shen \& Ho that Eddington ratio and orientation govern most of the diversity seen in broad-line quasar properties. 
\end{abstract}


\section{Introduction}\label{sec:intro}

The observed multi-wavelength properties of broad-line quasars display a remarkable regularity in the sense that these properties are covariant to a large extent, indicative of simple underlying physical mechanisms that are ultimately connected to the BH accretion process. The dominant trend is known as Eigenvector 1 (EV1), a physical sequence along which many quasar properties correlate with the strength of the optical \FeII\ emission. Since the discovery of EV1 by \citet{Boroson_Green_1992}, it has been frequently suggested that EV1 is mainly driven by the Eddington ratio ($\propto L/M$), based on the EV1 behaviors in terms of X-ray properties \citep[][]{Wang_etal_1996,Laor_1997}, \CIV\ properties \citep[][]{Wills_etal_2000,Sulentic_etal_2000a}, and later on more quantitative arguments based on the virial BH mass estimates that utilize the width of the broad emission lines \citep[][]{Laor_2000,Boroson_2002,Dong_etal_2011}.

A practical concern in the above arguments is the potential effect of orientation on the observed broad \hbeta\ width, and hence the virial BH mass estimates. There is now ample evidence (both theoretical arguments and observations) that the broad-line region has a flattened geometry \citep[e.g.,][and references therein]{Wills_Browne_1986, Runnoe_etal_2013, Gaskell_2009}. Hence it is important to disentangle physical effects (such as due to accretion) from geometrical effects (such as due to orientation) when studying quasars in the context of EV1. \citet{Shen_Ho_2014} proposed that, based on several novel statistical tests with the largest low-redshift quasar sample to date, the diversity of quasar properties in the 2-dimensional EV1 plane defined by optical \FeII\ strength ($x$ axis) and broad \hbeta\ FWHM ($y$ axis) is mainly determined by the Eddington ratio (traced by \FeII\ strength) and orientation (traced by broad \hbeta\ FWHM) at fixed quasar luminosity. 

The primary methodology used in \citet{Shen_Ho_2014} to argue that EV1 is driven by Eddington ratio is quasar clustering, which can differentiate low-mass and high-mass quasars in a statistical sense \citep[e.g.,][]{Shen_etal_2009}, without using virial BH mass estimates that are prone to orientation and other systematic uncertainties \citep[e.g.,][]{Shen_2013}. An alternative approach is to measure the stellar velocity dispersion ($\sigma_*$) of quasar hosts, and use the $M-\sigma_*$ relation \citep[e.g.][]{Gebhardt_etal_2000a,Ferrarese_Merritt_2000} to infer the BH mass in quasars independently from the clustering test. 

In this Letter we perform the latter exercise. Using a large spectroscopic quasar sample drawn from the SDSS DR7 \citep{DR7}, we were able to measure host $\sigma_*$ as functions of quasar properties (such as luminosity, \hbeta\ width, and \FeII\ strength) by performing host/quasar spectral decomposition. We found that the host velocity dispersions follow consistent EV1 patterns as expected from the empirical framework in \citet{Shen_Ho_2014}. In \S\ref{sec:data} we describe the data and our spectral measurements; we present the main results in \S\ref{sec:results} and summarize our findings in \S\ref{sec:disc}. We adopt a flat $\Lambda$CDM cosmology with $\Omega_0=0.3$ and $H_0=70\,{\rm km\,s^{-1}Mpc^{-1}}$ throughout. For ease of discussion, we will use the terms quasar and Active Galactic Nucleus (AGN) interchangeably. 

\section{Data and Spectral Measurements}\label{sec:data}

\begin{figure}
    \includegraphics[width=0.48\textwidth]{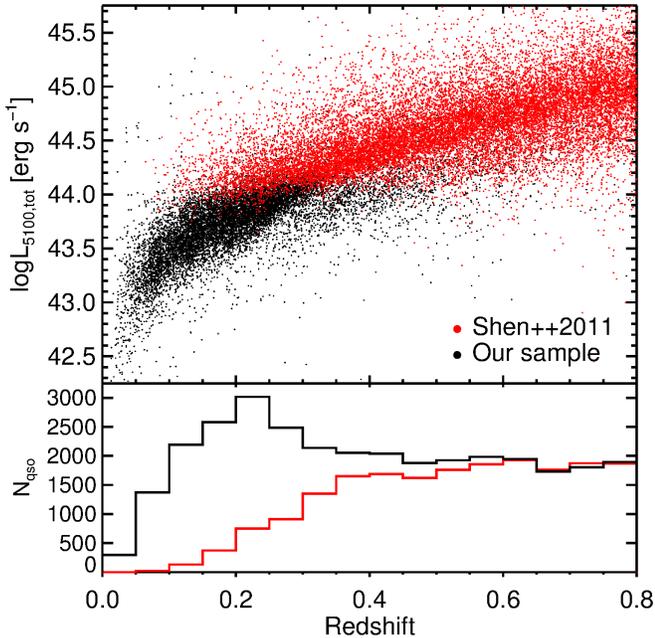}
    \caption{\textit{Top:} Distribution of quasars in the luminosity-redshift plane. The continuum luminosity is estimated at restframe 5100\,\AA\ from the full spectrum (before host decomposition). Objects included in the DR7 quasar catalog \citep{Shen_etal_2011} are shown in red dots, while the black dots show additional quasars included in our sample. The better coverage of our sample in the low-redshift (low-luminosity) region is clearly shown here. \textit{Bottom:} Redshift histograms of our sample (black) and the DR7 quasar catalog (red). }
    \label{fig:dist}
\end{figure}

Our quasar sample consists of all objects spectroscopically classified as broad-line quasars (\texttt{CLASS='QSO'}) from SDSS DR7 \citep{DR7}\footnote{http://das.sdss.org/www/html/}, with a redshift cut of $0.01<z<0.8$ such that we have a sufficient wavelength coverage to measure the host $\sigma_*$. 
Prior to the redshift cut, we have removed spectra from bad spectroscopic plates (PLATSN2=0), and we have checked the consistency between redshift values given in SDSS ``spSpec*'' files and those measured by the {\it idlspec2d} pipeline\footnote{Publicly available at http://www.sdss3.org/svn/repo/idlspec2d/}. Object spectra are discarded if both of the two redshifts are flagged as potentially problematic and inconsistent with each other. Our parent quasar sample contains 31346 unique objects. Fig.\ \ref{fig:dist} displays the redshift and luminosity distribution of our sample. Compared to the SDSS DR7 quasar catalog \citep{Schneider_etal_2010,Shen_etal_2011}, our sample expands the dynamic range in luminosity by including low-luminosity quasars that failed to meet the luminosity threshold in the DR7 quasar catalog. The wavelength coverage of SDSS spectra is 3800--9200\,\AA, with a spectral resolution of $R\sim 2000$, and a pixel size of $10^{-4}$ in $\log_{10}\lambda$, which corresponds to 69\,$\kms$. If there are duplicate spectra for the same quasar, the one with the higher S/N is adopted for spectral measurements. We do not distinguish radio-quiet and radio-loud quasars in this work, as the latter population is only $\sim 10\%$ of the entire quasar population \citep{Shen_etal_2011}. 

We follow \citet{vandenberk_etal_2006} and \citet{Shen_etal_2008b} to decompose the SDSS spectra into pure-host and pure-quasar spectra, using eigenspectra of quasars and galaxies constructed using the Principal Component Analysis (PCA) upon SDSS spectra \citep[][]{Yip_etal_2004a,Yip_etal_2004b}. This method assumes that the composite (quasar+galaxy) spectrum can be described by the combination of two independent sets of eigenspectra derived from pure galaxy and pure quasar samples. Empirically, \citet{vandenberk_etal_2006} found that a PCA decomposition using only the first few galaxy and quasar eigenspectra can reasonably recover the properties of the host galaxy, provided there is a significant contribution from the host galaxy in the composite spectrum. The spectral decomposition is deemed successful if the host fraction, $f_H=f_{\rm host}/(f_{\rm host}+f_{\rm qso})$, within $4160\,\textrm{\AA}<\lambda<4210\,$\AA\ is greater than 0.1 and less than 0.9. The lower boundary is found necessary in recovering the actual host galaxy properties  \citep[][]{vandenberk_etal_2006}, while the upper boundary is set to ensure successful measurements on quasar properties. We obtain the decomposed galaxy (quasar) spectrum by subtracting the reconstructed quasar (galaxy) spectrum from the original spectrum, and use the original spectral error arrays in the spectral measurements \citep[][]{vandenberk_etal_2006,Shen_etal_2008b}. We obtained successful decompositions for 20254 objects.

We then fit the decomposed host galaxy spectrum with the publicly available code \texttt{vdispfit} included in \textit{idlspec2d} to estimate the host $\sigma_*$. \texttt{vdispfit} is a template-based fitting method that performs the fits in the pixel space, where the templates consist of the first five PCA eigenspectra constructed from the echelle stellar spectra in \citet{Moultaka_etal_2004} and cover rest-frame wavelengths from 4125\,\AA\ to 6796\,\AA. During the fitting we mask narrow emission lines excited both by the galaxy and by the quasar, as well as the spectral region around the quasar broad \hbeta\ line (4760-5020\,\AA), where the host-quasar decomposition often shows significant residuals due to the small number of quasar eigenspectra used. We allow a small wavelength shift (within $\pm500\,{\rm km\,s^{-1}}$) in the fit to account for the uncertainty in the systemic redshift estimate. A fifth-order polynomial is also added to account for the broad-band continuum shape. The fit is restricted to the rest-frame wavelength range of 4125--5350\,\AA, which includes numerous stellar absorption features such as the G band (4304\,\AA), the Mg Ib $\lambda\lambda$ 5167,5173,5184 triplet, and Fe (5270\,\AA), but excludes the Ca H+K $\lambda\lambda$ 3969,3934 region. During the fit the templates are broadened to the native SDSS spectral resolution before the minimization, and the reported $\sigma_*$ values are already corrected for instrument broadening. It has been shown that $\sigma_*$ values measured by \texttt{vdispfit} are consistent with those measured by other fitting methods \cite[e.g.,][]{Shen_etal_2015}. We have also tried the empirical approach in \citet{Shen_etal_2008b} for aperture correction in $\sigma_*$, but found no significant changes in our results. Therefore we use the directly measured $\sigma_*$ in our following analysis.

We restrict our analysis to objects with well measured $\sigma_*$ by \texttt{vdispfit} ($\sigma_*/\Delta\sigma_* \geq 3$). Our final sample includes 9999 quasars with meaningful $\sigma_*$ measurements. We have compared our $\sigma_*$ measurements with those reported in \citet{Shen_etal_2008b} for the common objects, and found consistent results. The success of decomposing the host spectrum and measuring $\sigma_*$ depends strongly on the host-quasar contrast, which in turn introduces a luminosity-dependent selection bias such that more luminous quasars are less likely to yield a successful $\sigma_*$ measurement. We will return to this point in \S\ref{sec:results}. The typical S/N per SDSS pixel for objects with successful $\sigma_*$ measurements is $\sim 6$ in the decomposed host spectrum, and as a result many of our $\sigma_*$ measurements have large uncertainties (as much as $\sim 30\%$, possibly even larger given the systematic uncertainties). Nevertheless, the average $\sigma_*$ for binned objects should be well determined. Throughout this work we only use the average (estimated by the median) values of $\sigma_*$ in our binned analysis instead of individual measurements.

In addition to $\sigma_*$, we also measure the quasar spectral properties in the \hbeta\ region, including the continuum luminosity $L_{\rm 5100,AGN}\equiv \lambda L_{\lambda}$ at restframe 5100\,\AA, \OIII, \FeII\ and broad \hbeta\ properties, using the methodology outlined in earlier work \citep[e.g.,][]{Shen_etal_2008a,Shen_etal_2011}. These quantities are necessary to investigate the trends of $\sigma_*$ in terms of the EV1 correlations. Among those objects with meaningful $\sigma_*$ results, 7922 have good quasar measurements (e.g., the broad \hbeta\ line is detected at $>3\sigma$), which form the basis of our following analysis. We have visually inspected the objects that failed to produce good broad-line measurements but have good $\sigma_*$ measurements, and found that $\sim 60\%$ are due to pipeline mis-classifications (the object is a galaxy or a type 2 AGN) and the remaining are due to insufficient S/N or weak broad \hbeta\ in the fitting region (many of them show broad \halpha). The measurements for our final sample are provided in a supplemental online FITS table.




As a sanity check, our measurements reproduce the well known EV1 relations reported in previous work \citep{Boroson_Green_1992,Shen_Ho_2014}, and extend these results to fainter quasar luminosities enabled by our sample. 

\section{Results}\label{sec:results}

\begin{figure}
  \includegraphics[width=0.48\textwidth,angle=270]{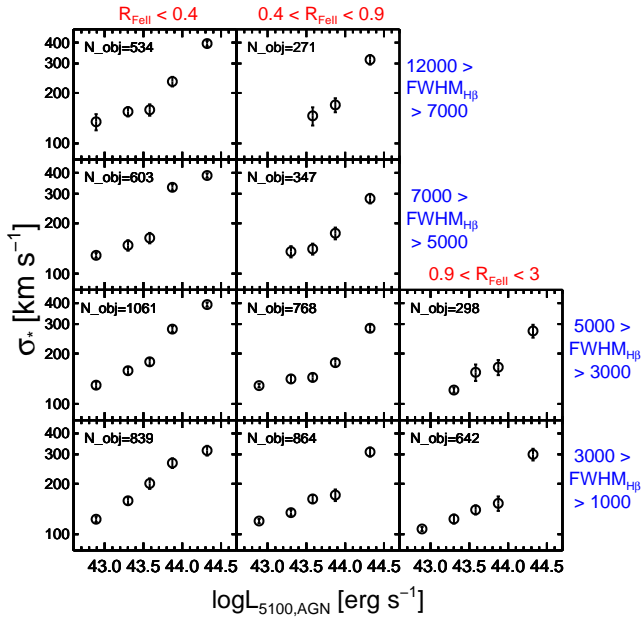}
  \caption{Median values of host-galaxy $\sigma_*$ in the EV1 plane. Columns correspond to increasing $\Rfe$ from left to right, and rows correspond to increasing broad \hbeta\ FWHM from bottom to top. The total number of objects in each $\Rfe-{\rm FWHM_{\rm H\beta}}$ bin is marked in each panel. The missing data points correspond to bins containing less than 50 objects. At fixed luminosity, there is a general decreasing trend of $\sigma_*$ with increasing $\Rfe$, suggesting that the Eddington ratio ($L/M$) increases with $\Rfe$. On the other hand, there is little dependence of $\sigma_*$ on the broad \hbeta\ FWHM at fixed $\Rfe$ and luminosity.}
  \label{fig:bin_sigma}
\end{figure}

We show our main result in Fig.\ \ref{fig:bin_sigma}. Our quasar sample with $\sigma_*$ and broad-line measurements was divided into subsamples in the two-dimensional EV1 plane of \FeII\ strength, measured by the ratio of \FeII\ EW within 4434--4684\,\AA\ to broad \hbeta\ EW ($\Rfe\equiv {\rm EW_{FeII}/EW_{H\beta}}$), and the FWHM of the broad \hbeta\ (${\rm FWHM}_{\rm H\beta}$). In each $\Rfe-{\rm FWHM_{\rm H\beta}}$ bin, we further divide the sample into different quasar luminosity bins. Statistically speaking, higher luminosity quasars are on average more massive BHs, hence we need to compare the $\sigma_*$ values at fixed quasar luminosity. 


\begin{figure}
  \centering
  \includegraphics[width=0.48\textwidth,angle=270]{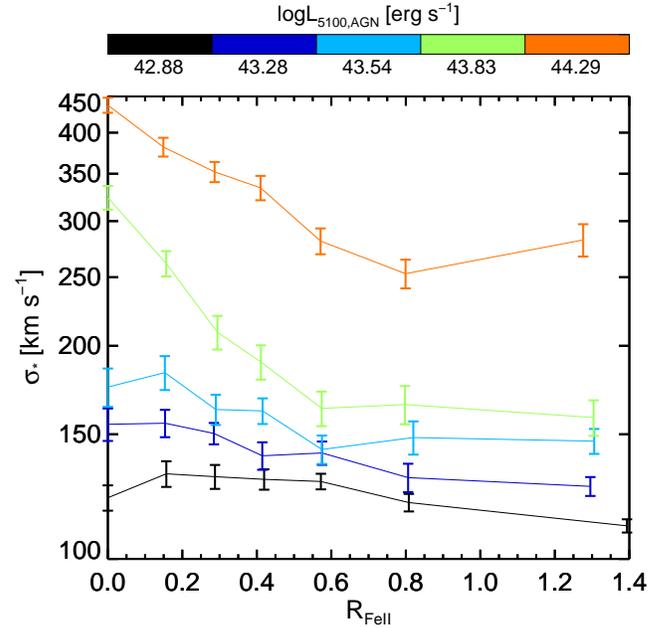}
  \caption{The median (with error) value of $\sigma_*$ as a function of $\Rfe$, for each luminosity bin shown in Fig.\ \ref{fig:bin_sigma}. Note that objects with different ${\rm FWHM_{H\beta}}$ are merged in those $L-\Rfe$ bins. A general trend of decreasing $\sigma_*$ with increasing $\Rfe$ is seen at all luminosities. The relatively less prominence in the trend at low luminosities is likely due to the selection bias against small $\sigma_*$ in our sample.}
  \label{fig:trend_sigma}
\end{figure}

Fig.\ \ref{fig:bin_sigma} demonstrates the behavior of $\sigma_*$ along the two axes of the 2D EV1 plane. In each ${\rm FWHM}_{\rm H\beta}$ row and at fixed quasar luminosity, there is a systematic trend in the average $\sigma_*$ as $\Rfe$ increases: quasars with high $\Rfe$ generally have lower host stellar velocity dispersion\footnote{\citet{Greene_Ho_2006} have shown that the $\sigma_*$ measurements could be biased in quasars with small-width broad \FeII\ emission when the fitting is restricted to certain spectral windows (typically several hundred \AA), particularly for high Eddington ratio objects (i.e., the high $\Rfe$ end of the EV1 sequence). Our $\sigma_*$ fitting is over a broad range of wavelengths, and the \FeII\ emission is largely removed with our host-quasar decomposition, both of which help remedy for this caveat. Even if we were fitting $\sigma_*$ similarly as \citet{Greene_Ho_2006}, for the ranges of Eddington ratios and $\sigma_*$ relevant in this work, the simulations in \citet{Greene_Ho_2006} suggest that any potential bias in $\sigma_*$ measurements is on the $\lesssim 10\%$ level, which is far too small to account for the trends in $\sigma_*$ seen in Figs.\ \ref{fig:bin_sigma} and \ref{fig:trend_sigma}. }. This suggests that the BH mass systematically decreases with $\Rfe$ when luminosity is fixed, hence the Eddington ratio increases with $\Rfe$. Another notable feature in Fig.\ \ref{fig:bin_sigma} is that there is little dependence of $\sigma_*$ on ${\rm FWHM_{H\beta}}$ when luminosity and $\Rfe$ are fixed, while the standard virial BH mass estimates predicts a strong vertical segregation in BH mass. The latter observation is consistent with the framework suggested by \citet{Shen_Ho_2014}: the dispersion in ${\rm FWHM_{H\beta}}$ at fixed luminosity and $\Rfe$ is largely due to orientation effects in a flattened BLR geometry, rather than intrinsic dispersion in BH mass. Therefore $\sigma_*$ remains more or less constant in different ${\rm FWHM_{H\beta}}$ bins when luminosity and $\Rfe$ are fixed. Given the insensitivity of $\sigma_*$ on ${\rm FWHM_{H\beta}}$ in the 2D EV1 plane, in the following analysis we only consider trends of $\sigma_*$ with luminosity and $\Rfe$ by merging objects with different ${\rm FWHM_{H\beta}}$ in each $L-\Rfe$ bin.

Fig.\ \ref{fig:trend_sigma} summarizes the dependence of host $\sigma_*$ as a function of $\Rfe$ at different quasar luminosities. Again, the general trend of decreasing average $\sigma_*$ with increasing $\Rfe$ is seen at all luminosities. However, one caveat is that the criterion to select well measured $\sigma_*$ (see \S\ref{sec:data}) and the SDSS spectral resolution may induce a bias against low $\sigma_*$ values. This may be responsible for the less prominent $\sigma_*-\Rfe$ trend seen for the low-luminosity quasars (with low $\sigma_*$ values) in Fig.\ \ref{fig:trend_sigma}.  

\begin{figure}[!b]
  \includegraphics[width=0.22\textwidth,angle=270]{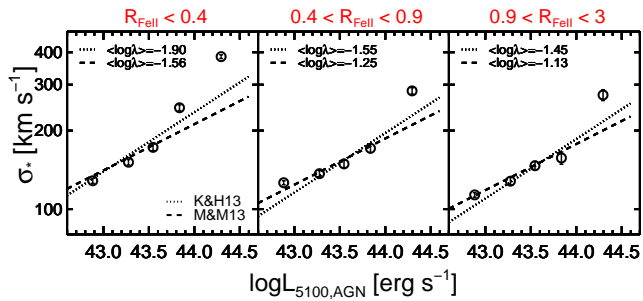}
  \caption{The median value of $\sigma_*$ as a function of luminosity in different ranges of $\Rfe$ (different panels). Objects with different ${\rm FWHM_{H\beta}}$ in a given $L-\Rfe$ bin are combined since there is little dependence of $\sigma_*$ on ${\rm FWHM_{H\beta}}$ when luminosity and $\Rfe$ are fixed (Fig.\ \ref{fig:bin_sigma}). In addition to our data points, we show simple relations expected from the observed local $M-\sigma_*$ relation, assuming a constant Eddington ratio $\bracket{\log\lambda}$ in each $\Rfe$ bin (e.g., Eqn.\ \ref{eqn:model}). The dotted and dashed lines represent the results using the observed $M-\sigma_*$ relation given by \citet{Kormendy_Ho_2013} and \citet{McConnell_Ma_2013}, respectively.}
  \label{fig:bin_comb}
\end{figure}

We point out that at fixed quasar luminosity, the fraction of quasars with measurable $\sigma_*$ does not show significant variation in each $\Rfe$-${\rm FWHM_{H\beta}}$ bin, hence there is no selection bias in the observed trend of $\sigma_*$ with $\Rfe$ or ${\rm FWHM_{H\beta}}$. On the other hand, $\sigma_*$ is generally more difficult to measure in higher-luminosity quasars, and those objects with measured $\sigma_*$ may be biased towards massive hosts (hence higher $\sigma_*$). In other words, the apparent trend of increasing $\sigma_*$ with luminosity may be subject to this luminosity-dependent selection bias in terms of $\sigma_*$ measurement. 

To assess the importance of this potential luminosity bias in $\sigma_*$ measurements, we make simple estimations using the observed $M-\sigma_*$ relation in local inactive galaxies \citep{Kormendy_Ho_2013,McConnell_Ma_2013}. Assuming a constant Eddington ratio $\lambda=L_{\rm bol}/L_{\rm Edd}$ in each $\Rfe$ bin (where $L_{\rm Edd}\equiv 1.26\times 10^{38}M/M_\odot\, {\rm erg\,s^{-1}}$ is the Eddington luminosity), we have an expected relation: 

\begin{equation}\label{eqn:model}
\log\left(\frac{\sigma_*}{200\,{\rm km\,s^{-1}}}\right)=\frac{1}{\beta}\left[\log \left(\frac{L_{\rm 5100,AGN}}{\rm erg\,s^{-1}}\right) - 37.1 - \alpha - \bracket{\log\lambda} \right]\ ,
\end{equation}
where we have assumed $\log L_{\rm bol}=10L_{5100,\rm AGN}$, and converted BH mass to $\sigma_*$ using the local $M-\sigma_*$ relation:
\begin{equation}
\log \left(\frac{M}{M_\odot}\right)=\alpha + \beta\log\left(\frac{\sigma_*}{{\rm 200\ km\,s^{-1}}}\right)\ ,
\end{equation} 
with $\alpha=8.49$ and $\beta=4.377$ as in \citet{Kormendy_Ho_2013}, and $\alpha=8.32$ and $\beta=5.64$ as in \citet{McConnell_Ma_2013}.

The predications are shown in Fig.\ \ref{fig:bin_comb} and compared with data. We simply adjusted the value of the average Eddington ratio $\bracket{\log\lambda}$ in each $\Rfe$ panel to best match the predictions with the data (as judged by eye). The data points for the highest luminosity bins significantly deviate from the predications, suggesting there might indeed be a strong selection effect in the highest luminosity bins that leads to overestimated $\sigma_*$ on average. In addition, as we mentioned earlier, the $\sigma_*$ measurements at the low-luminosity (low-$\sigma_*$) end may also be biased high due to our measurement quality cut and the SDSS spectral resolution, which will flatten the relations shown in Fig.\ \ref{fig:bin_comb}. Nevertheless, the general agreement in the trends between data and simple model predictions suggests that higher-luminosity quasars likely have intrinsically larger $\sigma_*$ (hence more massive BHs) on average when $\Rfe$ is fixed. This is consistent with the idea that $\Rfe$ is a good proxy for the Eddington ratio. As expected, the normalization of $\bracket{\log\lambda}$ required to match the predictions with data increases as $\Rfe$ increases; these Eddington ratio estimates, while subject to a number of caveats, are also reasonable values for the low-$z$ quasars in our sample. In a companion paper (Sun \& Shen, in preparation), we will explore the luminosity dependence of $\sigma_*$ more carefully with coadded spectra in the EV1 plane, which will allow us to better measure the average host velocity dispersion as a function of quasar luminosity that is less susceptible to the incompleteness of individual measurements.

\section{Discussion and Conclusions}\label{sec:disc}

We have studied the host stellar velocity dispersion in quasars as functions of quasar properties in the context of EV1, using a spectral PCA decomposition technique to separate the host galaxy and quasar components in integrated SDSS spectra. Our sample includes all low-redshift ($z<0.8$) quasars selected from the SDSS DR7, and extends to fainter luminosities than those included in the DR7 quasar catalog \citep{Schneider_etal_2010}. 

The trends of host stellar velocity dispersion $\sigma_*$ with EV1 properties revealed in this study provided new insights on the origin of the diversity of quasars in EV1 space. We found that at fixed quasar luminosity, $\sigma_*$ on average decreases with the optical \FeII\ strength, suggesting the average BH mass (Eddington ratio) decreases (increases) with \FeII\ strength, confirming the early suggestion that Eddington ratio drives the EV1 sequence \citep[e.g.,][]{Boroson_Green_1992,Boroson_2002}. On the other hand, at fixed quasar luminosity and \FeII\ strength, there is little dependence of $\sigma_*$ on the broad \hbeta\ FWHM, which is consistent with the suggestion in \citet{Shen_Ho_2014} that orientation plays a significant role in the dispersion in \hbeta\ FWHM at fixed \FeII\ strength. Although a quantitative mapping between \FeII\ strength and Eddington ratio is beyond the scope of this paper (which requires a more careful treatment of the selection effects and scrutiny on several additional assumptions), these findings on $\sigma_*$ provide further support to the empirical framework in \citet{Shen_Ho_2014} that Eddington ratio and orientation govern most of the diversity in quasar properties at fixed luminosity. 

Establishing Eddington ratio as the physical driver for the many covariant properties of quasars has important implications on the accretion physics of quasars. The change in accretion rate may modify the structure and the output spectrum of the accretion flow \citep[e.g.,][]{Wang_etal_2014}, and modulate the photoionization processes both in the nuclear region and in the more distant narrow line region, and therefore lead to the coherent patterns we observe as the main sequence of quasars. The empirical evidence presented here and in earlier work well motivates a theoretical investigation on the connection between quasar phenomenology and accretion processes. 

\acknowledgements  We thank the referee, Mike Brotherton, for useful comments that improved the manuscript, and Luis Ho, Todd Boroson and Charling Tao for helpful discussions. JS acknowledges support by the Department of Physics in Tsinghua University from the Tsinghua Xuetang (College) Program on Physics, and the hospitality of Carnegie Observatories during summer 2014. Support for the work of YS was provided by NASA through Hubble Fellowship grant number HST-HF-51314, awarded by the Space Telescope Science Institute, which is operated by the Association of Universities for Research in Astronomy, Inc., for NASA, under contract NAS 5-26555.  Funding for the SDSS and SDSS-II has been provided by the Alfred P. Sloan Foundation, the Participating Institutions, the National Science Foundation, the U.S. Department of Energy, the National Aeronautics and Space Administration, the Japanese Monbukagakusho, the Max Planck Society, and the Higher Education Funding Council for England. The SDSS Web Site is http://www.sdss.org/.


\end{document}